\documentclass[%
 reprint,
showpacs,preprintnumbers,
nofootinbib,
 amsmath,amssymb,
 aps,
 showkeys,
 showpacs
]{revtex4-1}
\usepackage{graphics}
\usepackage{bbold}
\usepackage{color}

\usepackage{cancel}
\usepackage[normalem]{ulem}

\usepackage{array}
\usepackage{booktabs}
\usepackage{amsmath}
\usepackage{graphicx}
\usepackage{subfigure}

\usepackage{amsmath}
\usepackage{slashed}

\usepackage{multirow}

\usepackage{float}

\usepackage{lineno,hyperref}

\newcommand{\division}{\mkern-\medmuskip\rotatebox[origin=c]{45}{\scalebox{0.9}{$-$}}\mkern-\medmuskip}

\def\noti	{\ensuremath{I\hspace{-1.7mm}\division}}

\def\be		{\begin{eqnarray}}
\def\en		{\end{eqnarray}}
\def\nen	{\nonumber\end{eqnarray}}
\def\no		{\nonumber}

\def\lt		{\left(}
\def\rt		{\right)}

\def\ds		{\ensuremath{\displaystyle}}

\def\ug		{\ensuremath{&=&}}
\def\jp		{\ensuremath{J\!/\psi}}

\def\ov		{\ensuremath{\overline}}
\def\ee		{\ensuremath{e^+e^-}}
\def\BB		{\ensuremath{B\bar{B}}}
\def\BBn		{\ensuremath{B^0\bar{B}^0}}

\def\ampl   {\ensuremath{\mathcal A}}
\def\De     {\ensuremath{D_e}}

\def\Fe     {\ensuremath{F_e}}

\def\LS		{\ensuremath{\Lambda \overline \Sigma{}^0}}
\def\LScc		{\ensuremath{\Lambda \overline \Sigma{}^0+{\rm c.c.}}}

\definecolor{amber}{rgb}{1.0, 0.55, 0.0}
\definecolor{redmod}{rgb}{0.89, 0.0, 0.13}
\definecolor{blumod}{rgb}{0.01, 0.28, 1.0}

\def \br {{\rm{BR}}}

\begin{document}

\title{
The cross section of $e^+e^- \to\LScc$ as a litmus test\\ of isospin violation\\ in the decays of vector charmonia into $\Lambda \overline \Sigma{}^0+{\rm c.c.}$.
}

\author{Rinaldo Baldini Ferroli}
\affiliation{%
Laboratori Nazionali di Frascati dell'INFN, Frascati, Italy}%
\author{Alessio Mangoni}
\affiliation{%
INFN Sezione di Perugia, I-06100, Perugia, Italy}%
\author{Simone Pacetti}
\affiliation{%
INFN Sezione di Perugia, I-06100, Perugia, Italy and Universit\`a di Perugia, I-06100, Perugia, Italy}%
\begin{abstract}
Under the aegis of isospin conservation, the amplitudes in Born approximation, i.e., considering the only one-photon-exchange mechanism, of the decay $\psi\to \Lambda \overline \Sigma{}^0+{\rm c.c.}$, where $\psi$ is a vector charmonium, and of the reaction $e^+e^- \to \Lambda \overline \Sigma{}^0+{\rm c.c.}$ at the $\psi$ mass, are parametrized by the same electromagnetic coupling. It follows that, the modulus of such a coupling can be extracted the data on the two observables: the decay branching fraction and the annihilation cross section.
\\
By considering the first two vector charmonia, $J/\psi$ and $\psi(2S)$, it is found that, especially in the case of $\psi(2S)$, there is a substantial discrepancy between the values of the modulus of the same electromagnetic coupling extracted from the branching ratio and the cross section.
\\
We propose, as a possible explanation for such a disagreement, the presence in the decay amplitude of an isospin-violating contribution driven by a mechanism based on physical (on-shell) intermediate states, that, due to their own nature, should be more effective in the $\psi(2S)$ decay than in that of the $J/\psi$.
\end{abstract}

\maketitle

\section{Introduction}
\label{sec:intro}
\begin{figure}[h]
\begin{center}
\includegraphics[width=0.8\columnwidth]{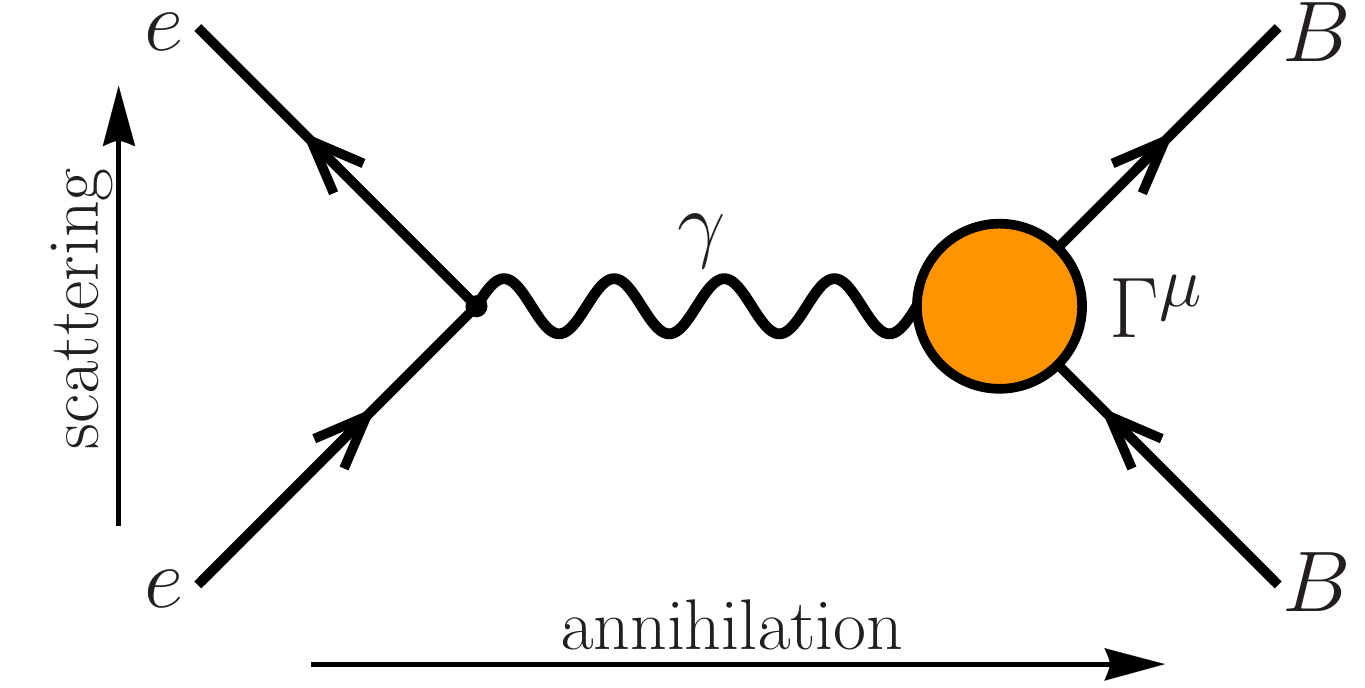}
\caption{\label{fig:annihi} Feynman diagram for the scattering $eB\to eB$ and the annihilations $\ee \leftrightarrow \BB$, in Born approximation. The non-constant $\Gamma^\mu$ matrix is associated to the vertex $BB\gamma$.}
	\end{center}
\end{figure}

In the framework of a Feynman-diagrammatic description, the electromagnetic form factors (FFs) are Lorentz scalar functions of $q^2$ associated with the vertex $hh\gamma$, where $q$ is the four-momentum of the photon and $h$ stands for a non-point-like hadron. The $q^2$-dependence of the FFs is a consequence of the finite spatial extension of the charge source of the interaction. Indeed, under suitable convergence conditions, the FFs can be interpreted as the Fourier transforms of the corresponding spatial charge densities. Therefore they play a fundamental role in the understanding of the dynamics and hence the structure of hadrons. 
\\
The hadrons studied in this work are the baryons belonging to the spin-$1/2$ SU(3) flavor-octet. The amplitude of the corresponding $B B \gamma$ vertex is described in terms of two independent FFs. Among the infinite possible choices, the pairs of FFs that are commonly used are:
\begin{itemize}
\item the so-called Dirac and Pauli FFs~\cite{Rosenbluth:1950yq}, $F_1^B(q^2)$ and $F_2^B(q^2)$, that parametrize the vector and tensor component of the amplitude;
\item the Sachs electric and magnetic FFs~\cite{Ernst:1960zza}, $G_E^B(q^2)$ and $G_M^B(q^2)$, that, in the reference frame where $q=(0,\vec{q})$, i.e., where there is no energy exchange, called Breit frame, represent the Fourier transforms of the electric and magnetic spatial densities of the baryon. 
\end{itemize}
As a consequence of the analyticity of the Feynman amplitudes, that can be proven order by order, the {\it physical} FFs are defined as the values for real arguments of functions that are analytic in the whole $z=q^2$-complex plane with a branch cut along the positive real axis, from the theoretical threshold $q^2=(2M_\pi)^2$ up to infinity, being $M_\pi$ the pion mass and the $\pi^+\pi^-$ state the lightest hadronic state that can couple with the $B\bar B$ final state.
\\
The Sachs FFs of the spin-1/2 baryons are experimentally accessible in both, space-like, $q^2<0$, and time-like region, $q^2>0$. In particular, in the space-like region their real values can be extracted from the Born differential cross section of the scattering process $eB\to eB$. While, in the time-like region, above the production threshold, $q^2=(2M_B)^2$, $M_B$ is the baryon mass, where FFs are complex, only their moduli can be measured. Their values are extracted from the differential cross sections of the annihilation processes $e^+e^-\leftrightarrow B\overline B$. The Feynman diagram of this reaction is shown in Fig.~\ref{fig:annihi}. By considering polarization observables, also the relative phase between the electric and magnetic Sachs FFs, $G_E^B$ and $G_M^B$, is measurable.
\section{Effective form factors and branching ratios}
The electromagnetic amplitude for the electron-positron annihilation into baryon-antibaryon
\be
e^-(p_1) e^+(p_2) \to \gamma^*(q) \to B(k_1) \overline B (k_2)\,,
\nen
where, in parentheses, are reported the four-momenta, is given by
\be
\mathcal{M}^\gamma_{\BB}=-{ie^2 \over q^2} \, \ov{v}(p_2)\gamma_\mu u(p_1) \, \ov{u}(k_1)\Gamma^\mu v(k_2) \,,
\nen
where $\Gamma^\mu$ is the most general four-vector, written in terms of the Dirac gamma matrices and the transferred four-momentum $q^\mu=p_1^\mu+p_2^\mu=k_1^\mu+k_2^\mu$, and parametrized in terms of Lorentz scalar functions of $q^2$, which represents the only non-constant scalar quantity that can be obtained from the four-momenta. Such scalar functions are the Dirac and Pauli FFs, $F_1^B$ and $F_2^B$, or equivalently, the Sachs electric and magnetic FFs $G_E^B$ and $G_M^B$. In the latter case, i.e., when $\Gamma^\mu$ is written in terms of the Sachs FFs, the expression, in Born approximation, for the total electromagnetic cross section of the annihilation $e^+ e^- \to B \bar B$ reads
\begin{equation}
\label{eq.csG}
\sigma_{\BB}(q^2) = {4 \pi \alpha^2\beta_{M_B}(q^2) \over 3q^2} \! \left(\! {2 M_B^2 \over q^2} \,|G_E^B(q^2)|^2 \!+\! |G_M^B(q^2)|^2 \!\!\right) \,,
\end{equation}
where $\sqrt{q^2}$ is the center of mass energy and
\be
\beta_{M_B}(q^2)=\sqrt{1- {4M_B^2 \over q^2}} \,,
\label{eq:velo}
\en
is the velocity of both out-going baryons in the \BB\ center-of-mass frame. We define the modulus of the electromagnetic effective FF 
\be
|\mathcal A^{\gamma}_{\BB}(q^2)|
 = \sqrt{|G_M^B(q^2)|^2+\frac{2 M_B^2 }{q^2}|G_E^B(q^2)|^2} \,,
\label{eq:abb}
\en
as the common modulus of the FFs multiplied by the kinematic factor $(1+2M_B^2/q^2)$, under the hypothesis 
\be
|G_M^B(q^2)|=|G_E^B(q^2)|\equiv |\mathcal A^{\gamma}_{\BB}(q^2) |\sqrt{1+\frac{2M_B^2}{q^2}}\,.
\nen 
Indeed, in terms of the effective FF, the expression of the total Born cross section of Eq.~\eqref{eq.csG} becomes
\begin{equation}
\label{eq.cs}
\sigma_{\BB} (q^2) = {4 \pi \alpha^2 \beta_{M_B}(q^2) \over 3q^2} |\mathcal A^{\gamma}_{\BB}(q^2)|^2 \,.
\end{equation}
We consider the decays of an SU(3) singlet meson $\psi$, i.e., a vector charmonium, and in particular, $\psi =J/\psi$, $\psi(2S)$, into pairs of spin-1/2 baryon-antibaryon $\BB$, belonging to the SU(3) octet, that can be represented by the matrix
\be
B=\begin{pmatrix}
\Lambda/\sqrt{6}+\Sigma^0/\sqrt{2} & \Sigma^+ & p\\
\Sigma^- & \Lambda/\sqrt{6}-\Sigma^0/\sqrt{2} & n\\
\Xi^- & \Xi^0 & -2 \Lambda/\sqrt{6}\end{pmatrix} \,.
\nen
The branching fraction (BR) of the pure electromagnetic (EM) decay 
\be
\psi\to\gamma^*\to \BB\,,
\nen
of the meson $\psi$, having mass $M_\psi$ and total width $\Gamma_\psi$, can be written in terms of the coupling constant $g^\psi_{\gamma}$ between the meson $\psi$ and the virtual photon $\gamma^*$, and the electromagnetic effective FF of Eq.~\eqref{eq:abb} at $q^2=M^2_\psi$ as it follows
\be
\label{eq.Brgen1}
\br_{\BB}^\gamma = {\left|g^\psi_{\gamma}\right|^2\beta_{M_B}(M_\psi^2) \over 16 \pi M_{\psi} \Gamma_{\psi}}  \,|\ampl^\gamma_{\BB}(M_\psi^2)|^2 \,.
\en
The modulus of the coupling constant $g^\psi_{\gamma}$ can be extracted by the BR for the EM decay $\psi\to\mu^+\mu^-$, that indeed has the expression 
\be
\br_{\mu^+\mu^-}^\gamma = {\left|g^\psi_{\gamma}\right|^2 \over 16 \pi M_{\psi} \Gamma_{\psi}}  \,,
\label{eq:mumu}
\en 
where the muon mass has been neglected.  
The expressions for the moduli of the combined amplitudes $g^\psi_\gamma\ampl^\gamma_{\BB}(M_\psi^2)$ in terms of two EM couplings $D_e$ and $F_e$, that parametrize the SU(3) symmetry breaking due to the EM interaction as defined in Ref.~\cite{Ferroli:2019nex} and references therein, are reported in Table~\ref{tab:A.par.gamma}. It is interesting to notice that, as a consequence of such a parametrization, we can define four ratios, either of cross section or of the corresponding EM BRs, that depend only on the masses of the baryons, i.e.,
\be
\frac{\sigma_{B_1\bar B_1}(M_\psi^2)}{\sigma_{B_2\bar B_2}(M_\psi^2)}
=\frac{\br^\gamma_{B_1\bar B_1}}{\br^\gamma_{B_2\bar B_2}}
\ug
\frac{\beta_{M_{B_1}}(M_\psi^2)}{\beta_{M_{B_2}}(M_\psi^2)}
\no\\\ug
\sqrt{\frac{M_\psi^2-4M_{B_1}^2}{M_\psi^2-4M_{B_2}^2}}\,,
\nen
with $\{B_1,B_2\}=\{n,\Xi^0\}$, $\{\Lambda,\Sigma^0\}$, $\{\Sigma^-,\Xi^-\}$, $\{p,\Sigma^+\}$. These ratios equal the unity in case of SU(3) symmetry restoration, i.e., in the limit of equal baryon masses, so that 
\be\frac{\sigma_{B_1\bar B_1}(M_\psi^2)}{\sigma_{B_2\bar B_2}(M_\psi^2)}
=\frac{\br^\gamma_{B_1\bar B_1}}{\br^\gamma_{B_2\bar B_2}}
=1+\mathcal O\lt M_{B_2}-M_{B_1}\rt\,, 
\nen
as $M_{B_2}\to M_{B_1}$.
\\
By using Eqs.~\eqref{eq.Brgen1} and~\eqref{eq:mumu}, the EM cross section $\sigma_{\BB}(M_\psi^2)$ of Eq.~\eqref{eq.cs}, for the annihilation process $ \ee \to \BB$ at $q^2=M_{\psi}^2$ can be expressed in terms of the EM BR of the decays $\psi \to\gamma^*\to \BB$ and $\psi \to\mu^+\mu^-$ as
\be
\sigma_{\BB} (M_{\psi}^2)= {\sigma^0_{\mu^+ \mu^-}(M_{\psi}^2) \over \br_{\mu^+ \mu^+}^\gamma} \, \br^\gamma_{ \BB} \,,
\label{eq.brsigma}
\en
where $\sigma^0_{\mu^+ \mu^-}(q^2)$ represents the bare, total Born cross section of the annihilation process $\ee \to \mu^+ \mu^-$, i.e.,
\be
\sigma^0_{\mu^+ \mu^-} (q^2)= {4 \pi \alpha^2 \over 3 q^2} \,.
\nen
In this case, as well as in that of the BR of Eq.~\eqref{eq:mumu}, the muon mass has been neglected, indeed the velocity of the outgoing muon is approximated to the unity and hence does not appear in the formula. On the other hand, in the ratio between the non-approximated expressions of the two quantities, namely, the cross section of the annihilation $\ee\to\mu^+\mu^-$ at the $\psi$ meson mass and the BR of the meson decay $\psi\to \mu^+\mu^-$, the muon velocity, that factorizes in both expressions, does cancel out. It follows that the cross section formula of Eq.~\eqref{eq.brsigma} is not affected by such an approximation.
\begin{table}[h]
\begin{center}
\caption{Parameterizations of the amplitudes of the EM decay $\psi\to\gamma^*\to\BB$ as function of the couplings $D_e$ and $F_e$~\cite{Ferroli:2019nex}.}
\label{tab:A.par.gamma} 
\begin{tabular}{lcl} 
\hline\noalign{\smallskip}
\BB &\hspace{10mm} & $g^\psi_\gamma\ampl^\gamma_{\BB}(M_\psi^2)$ \\
\noalign{\smallskip}\hline\noalign{\smallskip}
$\Sigma^0 \overline \Sigma{}^0$ && $\De$ \\
$\Lambda \overline \Lambda$ && $-\De$ \\
$\Lambda \overline \Sigma{}^0+ \rm{c.c.}$ && $\sqrt{3}\,\De$ \\
$p \overline p$ && $\De + \Fe$ \\
$n \overline n$ && $-2\,\De$ \\
$\Sigma^+ \overline \Sigma{}^-$ && $\De + \Fe$ \\
$\Sigma^- \overline \Sigma{}^+$ && $\De - \Fe$ \\
$\Xi^- \overline \Xi{}^+$ && $\De - \Fe$ \\
$\Xi^0 \overline \Xi{}^0$ && $-2\,\De$ \\
\noalign{\smallskip}\hline
\end{tabular}
\end{center}
\end{table}
\\
\section{Scaled cross sections from the branching ratios}
\label{sec:scaled}
The Feynman amplitude for the decay $\psi \to \Lambda \overline \Sigma{}^0 + \rm c.c.$, assuming isospin conservation, is purely EM~\cite{Ferroli:2019nex} and the BR, using the Eq.~\eqref{eq.Brgen1} and the amplitude parametrization reported in Table~\ref{tab:A.par.gamma}, can be written in terms of the modulus of the only EM coupling $D_e$ as
\be
\br_{\Lambda \overline \Sigma{}^0}^\gamma = {3 |\De|^2 \beta_{M_{\Lambda \overline \Sigma{}^0}}( M_{\psi}^2) \over 16 \pi M_{\psi} \Gamma_{\psi}} \,,
\label{eq.BRccDe}
\en
where $M_{\Lambda \overline \Sigma{}^0}(q^2)$ represents the mass term appearing in the cross section formula of Eq.~\eqref{eq.csG} for the pair $\Lambda\Sigma^0$ and it is given by
\be
M_{\Lambda \overline \Sigma{}^0}(q^2) = \sqrt{{1 \over 2}(M_{\Sigma^0}^2 + M_{\Lambda}^2) - {1 \over q^2}(M_{\Sigma^0}^2 - M_{\Lambda}^2)^2} \,.
\nen
The modulus of the EM coupling $\De$ can be extracted from the experimental value of the BR of the purely EM decay $\psi\to \Lambda \overline \Sigma{}^0  + \rm c.c.$, whose expression is given in Eq.~\eqref{eq.BRccDe}. In the cases of the $J/\psi$ and $\psi(2S)$ mesons, using the BRs reported on Table~\ref{tab:data}, we have
\begin{gather}
\begin{aligned}
J/\psi \ \ \rightarrow \ \ |\De| = (4.52 \pm 0.18) \times 10^{-4} \ \rm GeV \,, \\
\psi(2S) \ \ \rightarrow \ \ |\De| = (5.35 \pm 0.52) \times 10^{-4} \ \rm GeV \,, 
\label{eq.De}
\end{aligned}
\end{gather}
that are compatible with the values obtained, under similar hypotheses, in Refs.~\cite{Ferroli:2019nex,Ferroli:2020mra}. 
%
%
Since, just as that of the mixed state $\Lambda \overline \Sigma{}^0 + \rm c.c.$, all the EM amplitudes for the neutral final states, $B^0\bar B^0\in\mathcal N^0$, where
\be
\mathcal N^0\equiv \{ \LScc,n\bar n,\, \Lambda\bar\Lambda,\, \Sigma^0\bar\Sigma^0,\, \Xi^0\bar\Xi^0\}\,, 
\label{eq:N0-set}
\en
depend on the only EM coupling $D_e$, see Table~\ref{tab:A.par.gamma}, the BRs for the EM decays $\psi\to \gamma^*\to B^0\bar B^0$, as well as the $\ee\to B^0\bar B^0$ cross sections at the $\psi$ meson mass are proportional each other.\\ 
In particular, the BRs and the cross sections for the unmixed neutral final states can be expressed in terms of $\br_{\Lambda \overline \Sigma{}^0 }^\gamma$ of Eq.~\eqref{eq.BRccDe}, $\sigma_{\Lambda \overline \Sigma{}^0}(M_\psi^2)$ of Eqs.~\eqref{eq.cs} and~\eqref{eq.brsigma}, and the baryon velocity $\beta_{M_B}(q^2)$ of Eq.~\eqref{eq:velo}, as it follows
\be
\br_{B^0\bar B^0}^\gamma\ug \ds
\frac{N^2_{\BBn}\,\beta_{M_{B^0}}(M_{\psi}^2)}{\beta_{M_{\Lambda \overline \Sigma{}^0}}(M_\psi^2)}\,
\br_{\Lambda \overline \Sigma{}^0}^\gamma\,,\no
\\
&&\label{eq.BRemandS}
\\
\sigma_{B^0\bar B^0}(M_\psi^2)
\ug\ds
\frac{N^2_{\BBn}\,\beta_{M_{B^0}}(M_{\psi}^2)}{\beta_{M_{\Lambda \overline \Sigma{}^0}}(M_\psi^2)}\,
\sigma_{\Lambda \overline \Sigma{}^0 }(M_\psi^2)
\no\\
\ug
\ds
\frac{N^2_{\BBn}\,\beta_{M_{B^0}}(M_{\psi}^2)}{\beta_{M_{\Lambda \overline \Sigma{}^0}}(M_\psi^2)}\,
\frac{\sigma^0_{\mu^+\mu^-}(M_\psi^2)}{\br^\gamma_{\mu^+\mu^-}}\br_{\Lambda \overline \Sigma{}^0}^\gamma
\,,
\nen
with $\BBn\in\mathcal N^0$ and where the corresponding coefficients
\be
\label{eq.NB}
N_{\BBn} \equiv\left\{
\begin{array}{rcl}
1 && \BBn=\LScc\\
-2/\sqrt{3} && B_0= n\\
-1/\sqrt{3} && B_0= \Lambda\\
1/\sqrt{3} && B_0= \Sigma^0\\
-2/\sqrt{3} && B_0= \Xi^0\\
\end{array}\right.\,,
\en
are derived from Table~\ref{tab:A.par.gamma}.
\\
%
%
Using the values of $|\De|$ given in Eq.~\eqref{eq.De}, the EM BRs and the cross sections at the $J/\psi$ and $\psi(2S)$ masses can be computed by means of the expressions of Eq.~\eqref{eq.BRemandS}. The obtained results are reported in Table~\ref{tab:br.fromcc} and~\ref{tab:sigmas.fromcc}, respectively.
\begin{table} [H]
\vspace{-2mm}
\centering
\caption{Branching ratios data from PDG~\cite{Tanabashi:2018oca}.}
\label{tab:data} 
\begin{tabular}{lrr} 
\hline\noalign{\smallskip}
Decay process & Branching ratio & Error \\
\noalign{\smallskip}\hline\noalign{\smallskip}
$J/\psi \to \Lambda \overline \Sigma{}^0 + {\rm c.c.} $ & $(2.83 \pm 0.23) \times 10^{-5}$ & $8.13 \%$ \\
$\psi(2S) \to \Lambda \overline \Sigma{}^0 + {\rm c.c.} $ & $(1.23 \pm 0.24) \times 10^{-5}$ & $19.5 \%$ \\
$J/\psi \to \mu^+ \mu^- $ & $(5.961 \pm 0.033) \times 10^{-2}$ & $0.55 \%$ \\
$\psi(2S) \to \mu^+ \mu^- $ & $(8.0 \pm 0.8) \times 10^{-3}$ & $10 \%$ \\
\noalign{\smallskip}\hline
\end{tabular}
\end{table}%
\begin{table} [H]
\vspace{-2mm}
\centering
\caption{Electromagnetic BRs computed through Eq.~\eqref{eq.BRemandS} and using the values of $|\De|$ given in Eq.~\eqref{eq.De}.}
\label{tab:br.fromcc} 
\begin{tabular}{lrr} 
\hline\noalign{\smallskip}
Quantity & $\psi=J/\psi$ & $\psi=\psi(2S)$ \\
\noalign{\smallskip}\hline\noalign{\smallskip}
$\br^\gamma_{\Sigma^0 \overline \Sigma{}^0} $ & $(9.03 \pm 0.73) \times 10^{-6}$ & $(4.01 \pm 0.78) \times 10^{-6}$ \\
$\br^\gamma_{\Lambda \overline \Lambda} $ & $(9.82 \pm 0.80) \times 10^{-6}$ & $(4.19 \pm 0.82) \times 10^{-6}$ \\
$\br^\gamma_{n \overline n} $ & $(4.50 \pm 0.37) \times 10^{-5}$ & $(1.81 \pm 0.35) \times 10^{-5}$ \\
$\br^\gamma_{\Xi^0 \overline \Xi{}^0} $ & $(2.99 \pm 0.24) \times 10^{-5}$ & $(1.47 \pm 0.29) \times 10^{-5}$ \\
\noalign{\smallskip}\hline
\end{tabular}
\end{table}
\begin{table} [H]
\vspace{-2mm}
\centering
\caption{Electromagnetic cross sections computed through Eq.~\eqref{eq.BRemandS} and the values of $|\De|$ given in Eq.~\eqref{eq.De}.}
\label{tab:sigmas.fromcc} 
\begin{tabular}{l l  l} 
\hline\noalign{\smallskip}
Quantity & $q^2=M_{J/\psi}^2$ & $q^2=M_{\psi(2S)}^2$ \\
\noalign{\smallskip}\hline\noalign{\smallskip}
$\sigma_{\Sigma^0 \overline \Sigma{}^0}(q^2) $ & $(1.37 \pm 0.11) \ \rm pb$ & $(3.20 \pm 0.67) \ \rm pb$ \\
$\sigma_{\Lambda \overline \Lambda}(q^2) $ & $(1.49 \pm 0.12) \ \rm pb$ & $(3.35 \pm 0.70) \ \rm pb$ \\
$\sigma_{n \overline n}(q^2) $ & $(6.84 \pm 0.56) \ \rm pb$ & $(14.5 \pm 3.0) \ \rm pb$ \\
$\sigma_{\Xi^0 \overline \Xi{}^0}(q^2) $ & $(4.54 \pm 0.37) \ \rm pb$ & $(11.8 \pm 2.5) \ \rm pb$ \\
$\sigma_{\Lambda \overline \Sigma{}^0}(q^2)$ & $(4.30 \pm 0.35) \ \rm pb$ & $(9.8 \pm 2.1) \ \rm pb$ \\
\noalign{\smallskip}\hline
\end{tabular}
\end{table}
\noindent
We can compare the cross sections values of Table~\ref{tab:sigmas.fromcc} with those obtained by the direct measurements on the corresponding annihilation reactions $\ee\to\BBn$, performed at the \ee\ experiments BESIII~\cite{bes-nn} and BaBar~\cite{Aubert:2007uf}. 
By taking advantage from the general expressions of Eq.~\eqref{eq.BRemandS}, we define the \emph{scaled} cross section
\be
\tilde \sigma(q^2) \equiv \frac{\sigma_{B^0\bar B^0}(q^2)}{N_{\BBn}^2\,\beta_{M_{B^0}}(q^2)}
=\frac{4 \pi \alpha^2}{3q^2} |\mathcal A^{\gamma}_{\LS}(q^2)|^2\,,
\label{eq.tilde.sigma}
\en
where the last identity follows from the expression of the generic cross section $\sigma_{\BBn}(q^2)$ in terms of $\sigma_{\LS}(q^2)$, together with that of $\sigma_{\LS}(q^2)$ itself in terms of the modulus of the effective FF $\mathcal A^{\gamma}_{\LS}(q^2)$ of Eq.~\eqref{eq.cs}. It is clear that the scaled cross section does not dependent on the baryon pair. In particular, starting from the last member of Eq.~\eqref{eq.tilde.sigma}, at the $\psi$ meson mass, we have also the following expression
\be
\!\!\!\!\!
\tilde \sigma (M_{\psi}^2) =\frac{4 \pi \alpha^2}{3M_\psi^2} |\mathcal A^{\gamma}_{\LS}(M_\psi^2)|^2
= {\alpha^2 |\De|^2 \over 4 M_\psi^3 \Gamma_\psi \br_{\mu^+\mu^-}^\psi}\,,
\label{eq:xs-tilde-cc}
\en
where we have used $|\mathcal A^{\gamma}_{\LS}(M_\psi^2)|^2=3|\De|^2/|g_\gamma^\psi|^2$, for the modulus of the effective FF at the $\psi$ mass, see Table~\ref{tab:A.par.gamma}, and $|g^\psi_\gamma|^2=16\pi M_\psi\Gamma_\psi\,\br_{\mu^+\mu^-}^\gamma$, from Eq.~\eqref{eq:mumu}. The scaled cross sections at the $J/\psi$ and $\psi(2S)$ masses, corresponding to the scaled values of the individual cross sections reported in Table~\ref{tab:sigmas.fromcc}, are
\be
\begin{array}{rcl}
\tilde \sigma(M_{J/\psi}^2) \ug (6.45 \pm 0.54) \ {\rm pb} \,,  \\
\tilde \sigma(M_{\psi(2S)}^2) \ug (12.6 \pm 2.6) \ {\rm pb}  \,.
\end{array}
\label{eq:xs-tilde-br}
\en
\section{Direct measurement of scaled cross section}
\label{sec:scaled-ee}
A fitting procedure has been used in order to extract the experimental values of the scaled cross section at the $J/\psi$ and $\psi(2S)$ masses from the BESIII and BaBar data on the total cross section of the reactions $\ee\to B^0\bar B^0$. Being especially interested at the $J/\psi$ and $\psi(2S)$ mass region, we assume for the effective EM FF of Eq.~\eqref{eq:abb} the high-$q^2$-power-law behavior predicted by the perturbative QCD~\cite{Matveev:1973uz,Brodsky:1973kr}, i.e., 
\be
\mathcal A^{\gamma}_{\BB}(q^2)=\mathcal{O}\lt(q^2)^{-2}\rt\,, \ \ \ q^2\to \pm\infty\,.
\label{eq:asy}
\en
As a consequence, including also the logarithmic QCD correction, we define for the scaled cross section the fit function
\be
\tilde\sigma_{\rm fit}(q^2) = {A \over (q^{2})^{5} \lt\pi^2 + \ln^2( q^2 / \Lambda_{\rm QCD}^2 ) \rt^2 } \,,
\label{eq.fit}
\en
where $A$ is a dimensional\footnote{Dim$(A)=$ [energy]$^8$ in natural units: $\hbar=c=1$.} free parameter to be determined by a standard $\chi^2$ minimization procedure, while the QCD scale is kept fixed at $\Lambda_{\rm QCD} = 0.35 \, \rm GeV$.
\\
The fit has been performed on 
\be
N_{\rm Exp}=\sum_{\BBn\in\mathcal{N}^0}D_{\BBn}
\nen
 scaled cross section data points, obtained by the BaBar and the BESIII experiment, where $D_{\BBn}$ is the number data points on the specific reaction $\ee\to\BBn$, with $\BBn\in\mathcal{N}^0$. Moreover, to avoid the threshold energy regions, where, by definition, the cross sections do not follow the power-law behavior of Eq.~\eqref{eq:asy}, the cut-off value $q^2_{\rm asy}\equiv~(2.8\,{\rm GeV})^2$ has been defined and only data at $ q^2 \ge q^2_{\rm asy}$ have been considered.\\ 
 More in detail, the data set of the reactions $\ee\to\BBn$ is
\be
\left\{q_{\BBn,j}^2,\sigma_{\BBn}^{(j)},\delta\sigma_{\BBn}^{(j)}\right\}_{j=1}^{D_{\BBn}}\,, \ 
\ \  \BBn\in\mathcal N^0\,,
\nen
it contains $D_{\BBn}$ data points representing the cross section values $\sigma_{\BBn}^{(j)} \pm\delta\sigma_{\BBn}^{(j)}$ measured at  $q_{\BBn,j}^2\ge q^2_{\rm asy}$. The corresponding values of the scaled cross section, obtained through the expression of Eq.~\eqref{eq.tilde.sigma}, are
\be
\tilde\sigma^{(j)}_{\BBn}\ug  \frac{\sigma_{\BBn}^{(j)}}{N_{\BBn}^2\,\beta_{M_{B^0}}(q^2_{\BBn,j})}\,, \no\\ 
\delta\tilde\sigma^{(j)}_{\BBn}\ug \frac{\delta\sigma_{\BBn}^{(k,j)}}{N_{\BBn}^2\,\beta_{M_{B^0}}(q^2_{\BBn,j})}\,,
\nen
with $j=1,2,\ldots,D_{\BBn}$ and $\BBn\in\mathcal N^0$. It follows that 
all data points can be collected in the unique set
\be
\left\{\left\{q_{\BBn,j}^2,\tilde\sigma^{(j)}_{\BBn},\delta\tilde\sigma^{(j)}_{\BBn}\right\}_{j=1}^{D_{\BBn}}\right\}_{\BBn\in\mathcal{N}^0}\,.
\nen
\begin{figure}[H]
\includegraphics[width=0.95\columnwidth]{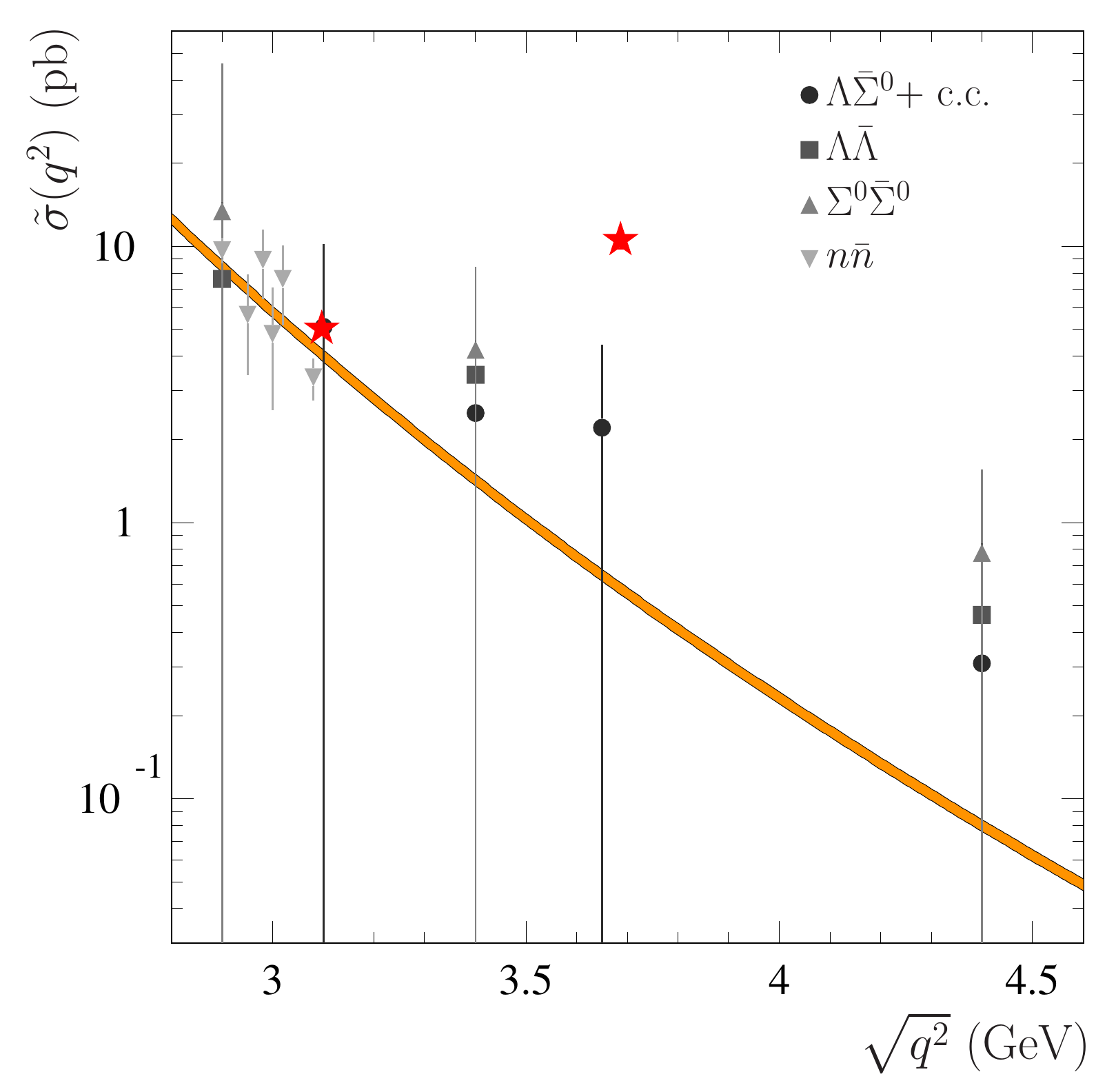}
\caption{\label{fig.data-fit-a} The solid points are the data on the scaled cross section from BESIII~\cite{bes-nn} and BABAR~\cite{Aubert:2007uf}, the orange band represents the fit results including the errors, while the red stars indicate the values of the scaled cross section at the $J/\psi$ and $\psi(2S)$ masses derived by the BRs of the corresponding decays $J/\psi\to\LScc$ and $\psi(2S)\to\LScc$.}
\end{figure}\noindent
The data on the scaled cross section that have been used are shown in Fig.~\ref{fig.data-fit-a} as solid points, together with the fit function, represented by the orange band that includes the error, and the values of the scaled cross section of Eq.~\eqref{eq:xs-tilde-br}, obtained by the BRs at the $J/\psi$ and $\psi(2S)$ masses, indicated by two red stars.
\\
The best value for the parameter $A$, that defines the fit function for the scaled cross section of Eq.~\eqref{eq.fit}, is 
\be
A=(3.29 \pm 0.27)\times 10^{8} \ \rm GeV^{10} \, pb\,.
\nen
It has been obtained by minimizing the $\chi^2$
\be
\chi^2=\sum_{\BBn\in\mathcal N^0}\sum_{j=1}^{D_{\BBn}}
\lt \frac{\tilde \sigma_{\rm fit}(q^2_{\BBn,j})-\tilde \sigma^{(j)}_{\BBn}}{\delta\tilde \sigma^{(j)}_{\BBn}}\rt^2\,,
\nen
 on four data sets for the scaled cross section, i.e., in this case $\mathcal N^0=\{ \LScc,n\bar n,\Lambda\bar\Lambda,\Sigma^0\bar\Sigma^0\}$, with a total of $D_{\rm Exp}=16$ data points, with the specific cardinalities: $D_{\LS}=4$, $N_{n\bar n}=6$, $D_{\Lambda\bar\Lambda}=3$, $D_{\Sigma^0\bar\Sigma^0}=3$.
 \\
 The predictions for the scaled cross section at the $J/\psi$ and $\psi(2S)$ masses, based on the $\ee\to\BBn$ cross section data of Table~\ref{tab:sigmas.fromcc}, are
  \be
  \begin{array}{rcl}
 \tilde\sigma_{\rm fit}(M_{J/\psi}^2)\ug (4.86  \pm 0.44)\,{\rm pb}\,,\\
 \tilde\sigma_{\rm fit}(M_{\psi(2S)}^2)\ug (0.692\pm 0.096)\,{\rm pb}\,,\\
 \end{array}
 \label{eq:xs-fit}	
 \en
and have to be compared with the corresponding values obtained by using the BRs of the decays $J/\psi\to\LScc$ and $\psi(2S)\to\LScc$, reported in Eq.~\eqref{eq:xs-tilde-br}, i.e.,
 \be
 \begin{array}{rcl}
\tilde \sigma(M_{J/\psi}^2) \ug (6.45 \pm 0.54) \ {\rm pb} \,,  \\
\tilde \sigma(M_{\psi(2S)}^2) \ug (12.6 \pm 2.6) \ {\rm pb}  \,.\\
\end{array}
 \nen
There is an evident discrepancy, especially for the value at the $\psi(2S)$ mass, where there is a difference of more than 4.6 standard deviations, while it is less than 2.3 at the $J/\psi$ mass. Moreover, it appears even more intriguing the increasing behavior with $q^2$ shown by the results of Eq.~\eqref{eq:xs-tilde-br}, where indeed a fast decreasing trend is expected.
\section{Reconciling the cross section and branching ratio data}
\label{sec:riconcilia}
\begin{figure}[h!]
\includegraphics[width=0.95\columnwidth]{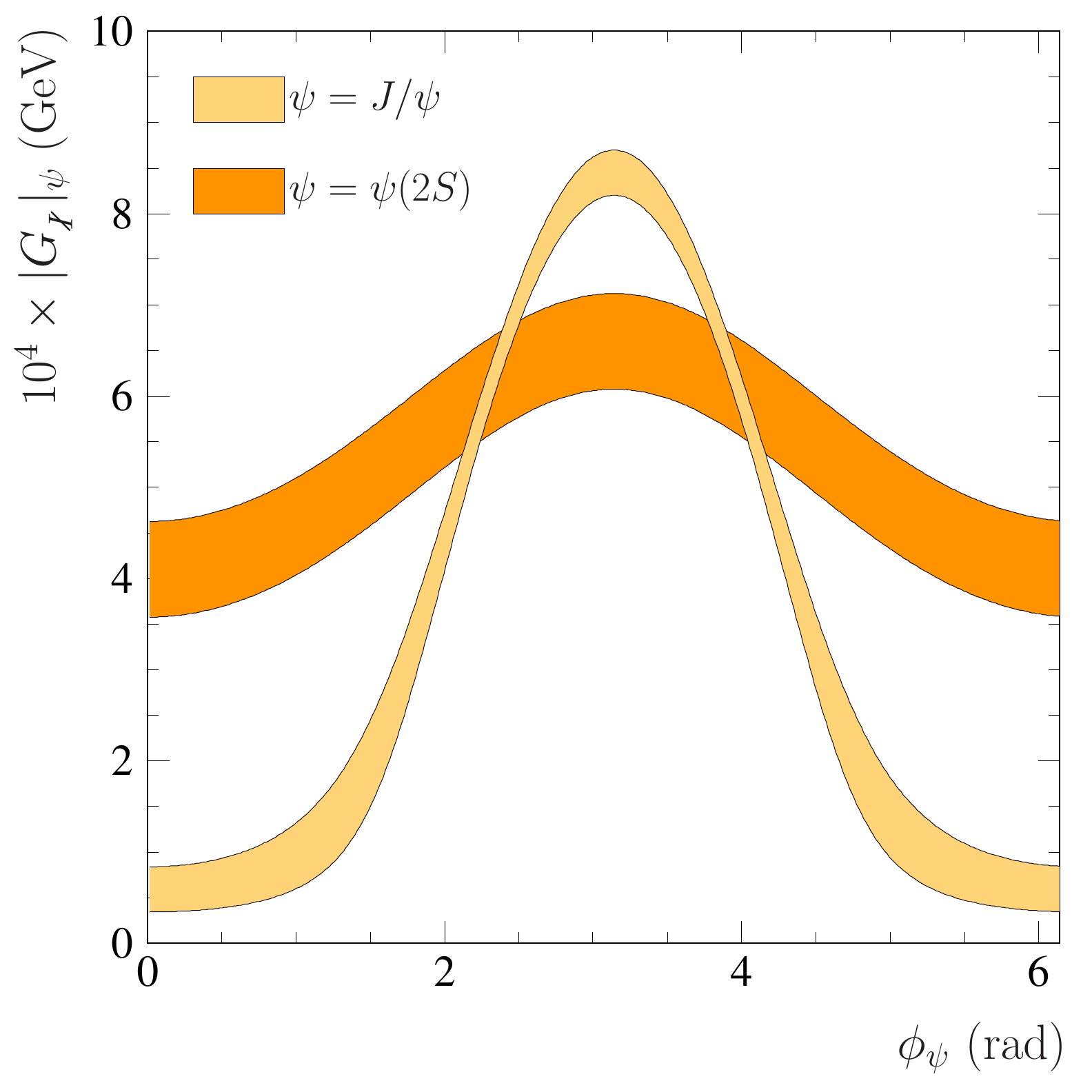}
\caption{\label{fig:iso-vio} Modulus of the isospin-violating amplitude of the $J/\psi$ meson, light-orange band, and of the $\psi(2S)$, dark-orange band, as a function of the phase relative to the EM amplitude \De.}
\end{figure}

\begin{figure}[h!]
  	\begin{center}
  		\includegraphics[height = 50 mm]{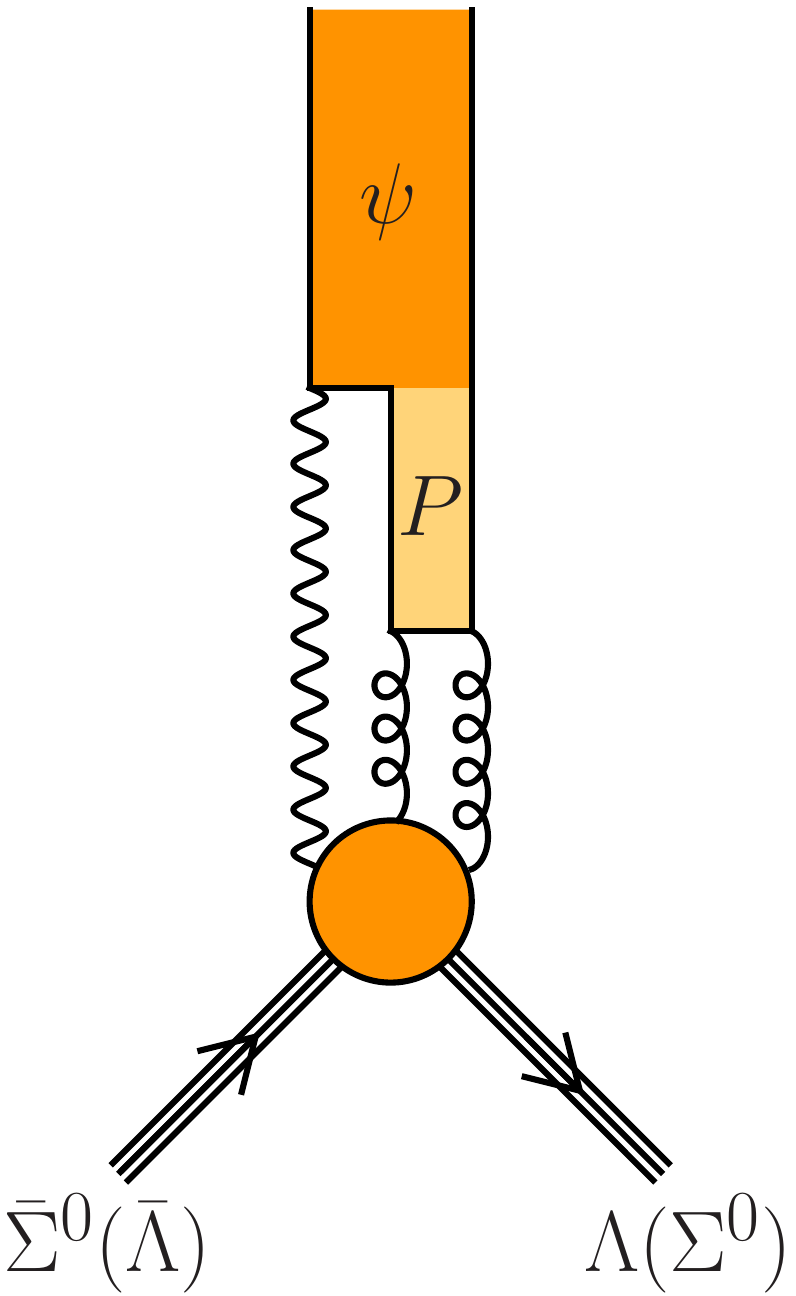}\hspace{10 mm}%
  		\includegraphics[height = 50 mm]{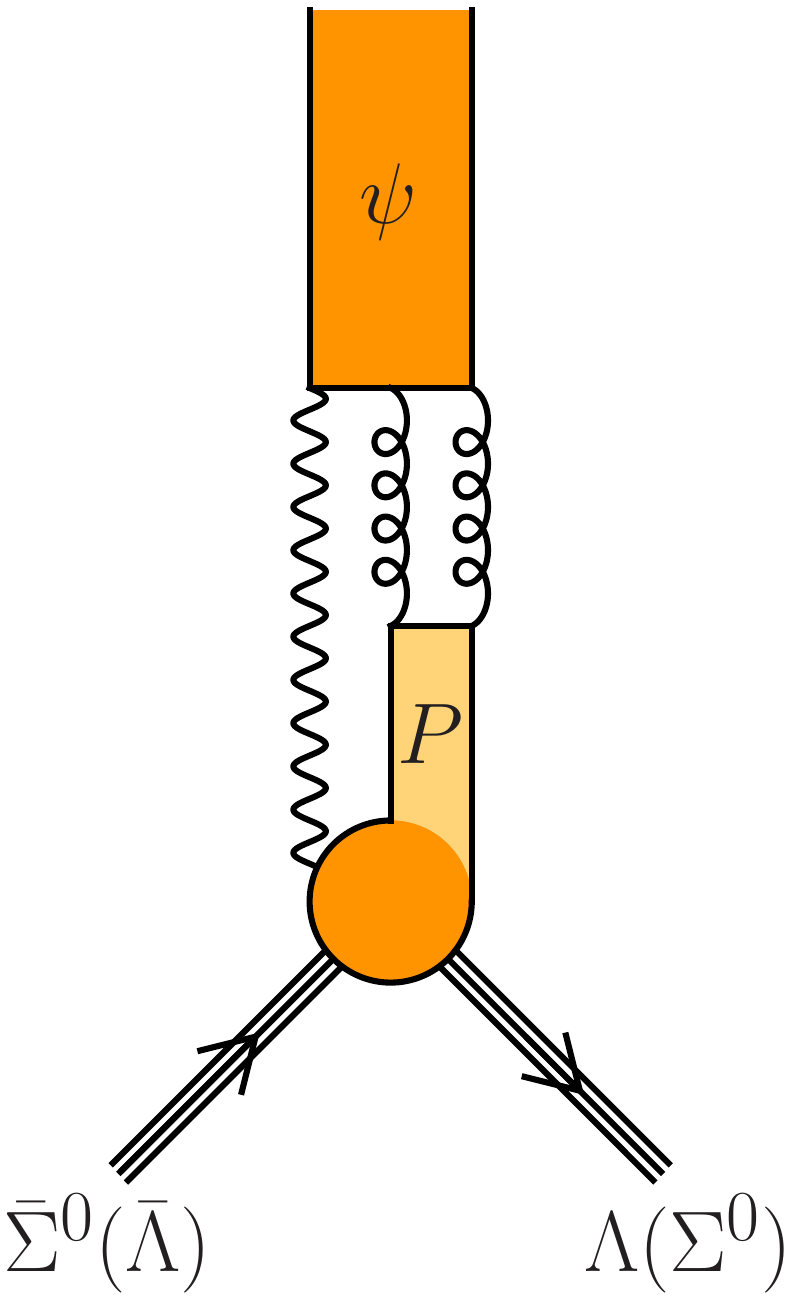}
		\caption{\label{fig:gg-gamma}Feynman diagrams of $\psi\to P\gamma \to gg\gamma\to\LScc$, left panel, and $\psi\to gg\gamma \to P\gamma \to\LScc$, right panel, isospin-violating contributions to the decay $\psi\to\LScc$.}
  	\end{center}
  \end{figure}

By assuming the parameterizations reported in  Table~\ref{tab:A.par.gamma} for the amplitudes of the EM decay $\psi\to\gamma^*\to\BB$, where $\psi$ stands for a vector charmonium, the modulus of the EM amplitude $\De$ can be extracted from the measurements of two different observables, the BRs of the decays $\psi\to\LScc$,  and the cross sections, at the $\psi$ mass, of the reactions $\ee\to\BBn$ for any neutral baryon pairs $\BBn$ belonging to the set $\mathcal N^0$ of Eq.~\eqref{eq:N0-set}.
\\
The level of agreement between the values of $|\De|$ obtained by these two independent experimental sources, BRs and cross sections, is the measure of the goodness of the hypotheses underlying the expressions of Eqs.~\eqref{eq.BRccDe} and~\eqref{eq:xs-tilde-cc}. 
\\
While the formula of Eq.~\eqref{eq:xs-tilde-cc}, that gives the Born cross section of the reaction $\ee\to\BBn$ in terms of $|\De|$ does not need any further assumption, besides the amplitude parameterization of Table~\ref{tab:A.par.gamma}, that of Eq.~\eqref{eq.BRccDe}, for the BR of the decay $\psi\to\LScc$, does require, instead, the crucial hypothesis of isospin conservation. Indeed, it is just under the aegis of isospin conservation that the decay $\psi\to\LScc$ proceeds only electromagnetically, through the one-photon exchange process (Born approximation) $\psi\to\gamma^*\to\LScc$.
\\
On the other hand, by allowing an isospin-violating contribution, $G_{\noti}$, to the decay amplitude, the expression  of Eq.~\eqref{eq.BRccDe} becomes 
\be
\br_{\LS}^{\gamma+\noti}\ug
\frac{3|\De+G_{\noti}|^2\beta_{\LS}(M_\psi^2)}{16\pi M_\psi\Gamma_\psi}
\no\\
\ug
\frac{3\lt |\De|^2+|G_{\noti}|^2 +2|\De||B_{\noti}| \cos(\phi)\rt\beta_{\LS}(M_\psi^2)}{16\pi M_\psi\Gamma_\psi}
\no\\
\ug
\br_{\LS}^{\gamma}+\br_{\LS}^{\noti}+\br^{\sim}_{\LS}\,,
\nen
where $\phi$ is the relative phase between the amplitudes $\De$ and $G_{\noti}$, i.e., $\phi=\arg(\De/G_{\noti})$, and moreover, we have single out the three BRs
\be
\br_{\LS}^{\gamma}
\ug 
\frac{3|\De|^2\beta_{\LS}(M_\psi^2)}{16\pi M_\psi\Gamma_\psi}\,,
\no
\\
\br_{\LS}^{\noti}
\ug
\frac{3|G_{\noti}|^2\beta_{\LS}(M_\psi^2)}{16\pi M_\psi\Gamma_\psi}
\,,
\no\\
\br^{\sim}_{\LS}\ug
\frac{6|\De||G_{\noti}|\cos(\phi)\beta_{\LS}(M_\psi^2)}{16\pi M_\psi\Gamma_\psi}\,,
\nen
as the EM, isospin-violating and interference contributions, respectively.
\\
In the light of this new interpretation, the values of $|\De|$ extracted from the BRs of the decays $J/\psi\to\LScc$ and $\psi(2S)\to\LScc$, given in Eq.~\eqref{eq.De}, represent indeed the moduli of the total amplitudes $\De+G_{\noti}$, while the moduli of the pure EM amplitudes are obtained by the scaled cross section values of Eq.~\eqref{eq:xs-fit} using the expression of Eq.~\eqref{eq:xs-tilde-cc}, in summary 
\begin{gather}
\label{eq.De.all}
\begin{aligned}
|\De+G_{\noti}|_{J/\psi} \ug (4.52 \pm 0.18) \times 10^{-4} \ {\rm GeV} \,, \\
|\De|_{J/\psi} \ug (3.93 \pm 0.17) \times 10^{-4} \ {\rm GeV}\,, \\
|\De+G_{\noti}|_{\psi(2S)} \ug (5.35 \pm 0.52) \times 10^{-4} \ {\rm GeV} \,, \\
|\De|_{\psi(2S)} \ug (1.25 \pm 0.07) \times 10^{-4} \ {\rm GeV} \,.
\end{aligned}
\end{gather}
The moduli of the EM and isospin-violating amplitudes can be obtained as functions of the relative phases, i.e., 
\be
|G_{\noti}|_\psi = \sqrt{|\De\!+\!G_{\noti}|_\psi^2\! -\!|\De|_\psi^2 \sin^2(\phi_\psi)} \!-\! |\De|_\psi \cos (\phi_\psi) \,,
\nen
with $\psi=J/\psi$, $\psi(2S)$. Figure~\ref{fig:iso-vio} shows these moduli, the light-orange band for $J/\psi$ and the dark-orange for the $\psi(2S)$, the band width indicates the error. It is interesting to notice that the domain of the modulus $|G_{\noti}|_{\jp}$, as a function of the relative phase $\phi_{\jp}$, contains the one of $|G_{\noti}|_{\psi(2S)}$, in fact they are
\be
&&\ds (5.9\pm2.5)\!\cdot\! 10^{-5} <\frac{|G_{\noti}|_{J/\psi}}{{\rm GeV}}<(8.45\pm 0.25)\!\cdot \!10^{-4} \,,
\no\\
&&\ds (4.10\pm 0.52)\!\cdot\! 10^{-4}<\frac{|G_{\noti}|_{\psi(2S)}}{{\rm GeV}}< (6.60\pm 0.52)\!\cdot\! 10^{-4} \,.
\nen
Despite the fact that the maximum value of $|G_{\noti}|_{J/\psi}$, attained at $\phi_{J/\psi}=\pi$, is larger than the maximum of $|G_{\noti}|_{\psi(2S)}$, the most noticeable result is that the minimum of the latter modulus is significantly large. Such an eventuality implies that, in the case of the $\psi(2S)$ meson, the isospin-violating contribution, whatever its nature, is phenomenologically required.
\\
Indeed, while in the $J/\psi$ case, the minimum of $|G_{\noti}|_{J/\psi}$ is compatible with zero within about 2.6 sigmas, in the case of $\psi(2S)$, the minimum of $|G_{\noti}|_{\psi(2S)}$ is about 7.9 sigmas away from zero.
\\
 It follows  that whether an isospin-violating mechanism does contribute to the decay amplitude of a vector charmonium $\psi$ into the $\LScc$, such a mechanism is more effective in the case of the $\psi(2S)$ meson.
\\
A reason for this difference could be identified by invoking as a possible source of isospin-violation the presence of a gluon-gluon-photon, $gg\gamma$, intermediate state~\cite{Ferroli:2016jri}, besides the one-photon exchange mechanism. Following Ref.~\cite{Ferroli:2016jri}, the $gg\gamma$ amplitude could be described in terms of physical, i.e., on-shell, $P\gamma$ intermediate states, that produce the final state $\LScc$, where $P$ stands for either a pseudo-scalar or a tensor meson.
\\
If $P$ is a $c\bar c$ meson, it couples strongly to the vector charmonium $\psi$ and then decays into the gluon pair, the corresponding Feynman diagram is shown in the left panel of Fig.~\ref{fig:gg-gamma}. 	
\\
  \\
Instead, if $P$ is a light meson, it is produced by the hadronization of the gluon pair and then, by absorbing the photon, it produces the final state \LScc. The Feynman diagram of this reaction is shown in the right panel of Fig.~\ref{fig:gg-gamma}.
\\
  Mesons that can contribute to this $P\gamma \leftrightarrow gg\gamma$ intermediate states, having BRs larger than $10^{-3}$, together with the BRs themselves are listed in Table~\ref{tab:gggamma}. These BRs sum up to $\sim3\%$ of the total width of the $J/\psi$ meson, to be compared to $\sim 20\%$ of the total width of $\psi(2S)$. Such a quite clear hierarchy, assuming the model of Ref.~\cite{Ferroli:2016jri}, could be a possible explanation for the dominance of the isospin-violation amplitude in the $\psi(2S)$ decay with respect that of the $J/\psi$, i.e., $|G_{\noti}|_{\psi(2S)}\gg |G_{\noti}|_{J/\psi}$ in our parametrization.
\begin{table}[h]
\caption{\label{tab:gggamma}All available BRs larger than $10^{-3}$ for the decays $\psi\to P\gamma$, with $\psi=J/\psi$ and $\psi(2S)$~\cite{Tanabashi:2018oca}. The BR of the decay $J/\psi\to\eta(1405/1475)\gamma$ is the sum of those of 5 sub-channels.}
\begin{tabular}{l|l|l}\hline
	$\psi$ 
	& $P$ & $\br(\psi\to P\gamma)$ \\
	\hline
	\multirow{5}{*}{$J/\psi$} & $\eta_c(1S)$ & $(1.7\pm0.4)\%$\\
& $\eta'(958)$ & $(5.25\pm 0.07)\cdot 10^{-3}$\\
& $\eta(1405/1475)$ & $\sim 4.9\cdot 10^{-3}$\\
& $f_2(1270)$ & $(1.64\pm0.12)\cdot 10^{-3}$\\
& $\eta$ & $(1.108\pm0.027)\cdot 10^{-3}$\\
\hline
\multirow{3}{*}{$\psi(2S)$} 
& $\chi_{c0}(1P)$ & $(9.79\pm0.20)\%$\\
& $\chi_{c2}(1P)$ & $(9.52\pm0.20)\%$\\
& $\eta_{c}(1S)$ & $(3.4\pm0.5)\cdot 10^{-3}$\\
\end{tabular}  	
\end{table}
\section{Conclusions}
\label{sec:conclu}
Assuming isospin conservation, the decay $\psi\to\LScc$, where $\psi$ is a vector charmonium, proceeds only through the one-photon exchange mechanism. It follows that, in the framework of the parameterization described in Table~\ref{tab:A.par.gamma}, the modulus of corresponding unique EM amplitude \De\ can be extracted from the BR through the expression of Eq.~\eqref{eq.BRccDe}. The same quantity can be also obtained by measuring the cross section of the reaction $\ee\to\LScc$ at the $\psi$ mass, by taking advantage from the cross section formula of Eq.~\eqref{eq:xs-tilde-cc}.
\\ 
The degree of agreement between these two sources of experimental information on the same quantity, namely the modulus $|\De|$, measures the reliability of the hypotheses underlying the parameterizations, which relate $|\De|$ itself to the experimental observables. 
\\
By studying the $J/\psi$ and $\psi(2S)$ charmonia, it has been found that, while in the former case the moduli of the amplitude  $\De$ from the BR and the cross section are compatible within about 2.6 sigmas, in the case of $\psi(2S)$ there is instead a substantial disagreement, about 7.9 sigmas.
\\
A possible explanation for such a disagreement has been proposed and qualitatively argued in Sec.~\ref{sec:riconcilia}. In particular, this discordance has been ascribed to the presence of an isospin-violating contribution in the $\psi(2S)$ decay. 
\\
Another scenario could be also taken into account, i.e., the possibility of not complete reliability of the only available datum on $\br^\gamma_{\LS}$~\cite{Dobbs:2017hyd}. However, we do not consider seriously such an eventuality, because it should imply an overestimate of the BR by more than a factor of $18\pm4$, as can be deduced comparing the values reported in Tables~\ref{tab:data} and~\ref{tab:confronto}.
\\
Nevertheless, a new measurement, feasible at the $\tau$-charm factories, such as e.g. BESIII~\cite{Jiao:2016syk}, of the BR of the decay $\psi(2S)\to\LScc$ would be clarifying by adding crucial pieces of information on the eventual isospin violating contribution. 
\begin{table} [H]
\vspace{-2mm}
\centering
\caption{Electromagnetic BRs computed through Eq.~\eqref{eq.BRemandS} and using the values of $|\De|$ given in Eq.~\eqref{eq.De.all} (second and fourth equations).}
\label{tab:confronto} 
\begin{tabular}{lrr} 
\hline\noalign{\smallskip}
Quantity & $\psi=J/\psi$ & $\psi=\psi(2S)$ \\
\noalign{\smallskip}\hline\noalign{\smallskip}
$\br^\gamma_{\Sigma^0 \overline \Sigma{}^0} $ & $(6.81 \pm 0.61) \times 10^{-6}$ & $(2.20 \pm 0.26) \times 10^{-7}$ \\
$\br^\gamma_{\Lambda \overline \Lambda} $ & $(7.40 \pm 0.66) \times 10^{-6}$ & $(2.30 \pm 0.27) \times 10^{-7}$ \\
$\br^\gamma_{n \overline n} $ & $(3.39 \pm 0.30) \times 10^{-5}$ & $(9.9 \pm 1.1) \times 10^{-7}$ \\
$\br^\gamma_{\Xi^0 \overline \Xi{}^0} $ & $(2.25 \pm 0.20) \times 10^{-5}$ & $(8.10 \pm 0.95) \times 10^{-7}$ \\
$\br^\gamma_{\Lambda \overline \Sigma{}^0} $ & $(2.13 \pm 0.19) \times 10^{-5}$ & $(6.75 \pm 0.79) \times 10^{-7}$ \\
\noalign{\smallskip}\hline
\end{tabular}
\end{table}
\section*{Acknowledgement}
We would like to warmly acknowledge the Italian group of the {BESIII} Collaboration for the useful and fruitful discussions on experimental and phenomenological aspects concerning \jp\ decays.\\
This work was supported in part by the STRONG-2020 project of the European Union's Horizon 2020 research and innovation programme under grant agreement No.~824093.
%
%

%

\end{document}